# Making FORS2 fit for exoplanet observations (again)

H.M.J. Boffin, G. Blanchard, O.A. Gonzalez, S. Moehler, E. Sedaghati, N. Gibson, M.E. van den Ancker, J. Smoker, J. Anderson, C. Hummel, D. Dobrzycka, A. Smette, G. Rupprecht

ESO

For about three years, it was known that precision spectrophotometry with FORS2 suffered from systematic errors that made quantitative observations of planetary transits impossible. We identified the Longitudinal Atmospheric Dispersion Compensator (LADC) as the most likely culprit, and therefore engaged in a project to exchange the LADC prisms with the uncoated ones from FORS1. This led to a significant improvement in the depth of FORS2 zero points, a reduction in the systematic noise, and should make FORS2 again competitive for transmission spectroscopy of exoplanets.

Over the last two decades remarkable progress has been made in understanding the diversity of planets in our Galaxy from the success of radial velocity and transit surveys. Transiting planets allow both their masses (if radial velocities are also available) and radii to be measured leading to bulk densities and compositions. However, to truly understand planetary systems we need a method to obtain spectra of exoplanets, thereby probing the composition and structure of their atmospheres. Luckily, transiting systems allow such measurements despite being unable to spatially resolve the star and planet; rather we can temporally resolve light from the star and planet during transits, when the planet passes in front of its host.

The transit depth obtained by transmission spectroscopy of the host star provides a direct measurement of the planet-to-star radius ratio as a function of wavelength. The effective size of the planet varies due to wavelength dependent opacities in the planet upper atmosphere, and a transmission spectrum can therefore probe the atomic and molecular species in its atmosphere causing such radius variations (Seager & Sasselov, 2000; Brown, 2001; for a recent review, see Burrows 2014). For many years, space-based observations with HST and Spitzer were the only source of exoplanet spectra. This all changed after pioneering observations of the exoplanet GJ 1214b using FORS2 proved that precise transmission spectra are obtainable from ground-based instruments (Bean et al., 2010). Their paper used the FORS2 multi-object spectroscopy capability with the Mask Exchange Unit (MXU) to perform differential spectro-photometry, i.e. observing time-series spectra of the target star simultaneously with many comparison stars, thus correcting for variations in the Earth's atmospheric throughput. Since then, such observations have been performed routinely using different instruments and telescopes, and have even proved successful at infrared wavelengths (e.g., Snellen et al., 2010; Bean et al., 2011; Gibson et al., 2013a,b; Crossfield et al., 2013; Stevenson et al., 2013; Schlawin et al., 2014). As future spaced-based observatories will focus on the infrared, ground-based instrumentation will be the only way to probe the optical transmission spectra of exoplanets. This is a crucial wavelength regime to understand the physics of exoplanet atmospheres, in particular to determine the mean molecular weight of the atmosphere and atmospheric scale height from measuring the Rayleigh scattering slope.

A Cause for Alarm

Despite the pioneering observations by Bean et al. (2010; re-analysed in Bean et al. 2011), no other exoplanet transit has been published so far from FORS2 data and it seems that there is a consensus in the community that FORS2 is not suited for the study of planetary transits[1]. This is of course worrying, especially as the GMOS instruments on the Gemini telescopes, which are rather similar to FORS2, are among the most successful instruments to date for measuring transmission spectra from the ground (Gibson et al., 2013a,b; Crossfield et al., 2013).

The reason for this problem with FORS2 appears to lie in the unexpectedly high systematics in the differential light curves that were obtained, and that turned out to be impossible to calibrate out reliably (at least to the level of $\sim 10^{-4}$ as required in the transit depth precision for a hot Jupiter; see Fig. 1). The main source of these instrumental systematics is most likely the LADC on FORS2. From visual inspection, the anti-reflection coating of this optical element is indeed known to have degraded over time. Berta et al. (2011) report these systematics and mention: "Moehler et al. (2010) found that the LADC on the telescope has surface features that affect its sensitivity across the field of view. Because the LADC is positioned before the field rotator in the optical path and rotates relative to the sky, individual stars can drift across these features and encounter throughput variations that are not seen by the other comparison stars."

The design of the FORS2 LADC consists of two prisms of opposite orientation that are moved linearly with respect to each other, between 30 mm (park position) and 1100 mm. The forward prism does the dispersion correction, while the second prism corrects the pupil tilt, so that what remains is a variable image shift depending on the distance between the two prisms (Avila et al. 1997).

Exchanging the Prisms

The $MgF_2$ antireflective coatings of the FORS2 linear atmospheric dispersion corrector prisms have degraded since 1999, following an attempt to clean them. They show a lot of scattering (see Fig. 2). Since then, various cleanings have been performed to remove dust and paint (coming off the flat field screen), leading to further degradation. This degradation could be the cause of the systematics seen in the FORS2 transit data.

Indeed, a damaged coating may introduce:

- transmission loss larger than an uncoated set of prisms;
- scattering, leading to a decreased signal-to-noise ratio on any photometric measurements;
- variability of the transmission because of change in the humidity level.

A project was started in Paranal to address this issue. One aspect of it was to take advantage of the availability of spare parts from the twin instrument FORS1, which is now decommissioned. We therefore decided to uncoat the prisms of the FORS1 LADC and exchanged them with those previously in place in the FORS2 LADC. The removal of the damaged coating from the two prisms of the spare LADC was done by one of us (GBl; Fig. 3), using tools made of polyurethane and using Cerium oxyde (Opaline) for the polishing. The prisms of the LADC were then exchanged on 10 November 2014, while UT1 was undergoing maintenance.

---

[1] The fact that Bean et al. were successful in their observations is thought to be a combination of factors, i.e., short transit time, small variations in airmass during the observations, good weather, and more generally, the best of luck and great skills.

A battery of tests

A set of test observations were performed according to a commissioning plan before (28 to 30 October) and after (12 to 15 November) the prisms exchange. It is important to note that the coating of the primary mirror of UT1 had not been touched in the mean time, so that any change detected should be due to the LADC exchange.

These tests have allowed us to conclude that the exchanged prisms did not affect the image quality of the instrument and confirmed that the LADC was still efficient at correcting the atmospheric dispersion up to an airmass of 1.6. On the other hand, the uncoated prisms led to an increase in the measured zero points (Fig. 4): the implied gain in throughput is 0.12 (B filter), 0.08 (V), 0.06 (R), and 0.05 (I) magnitudes. This was also confirmed by measurement of spectrophotometric standard stars before and after the prism exchange. The gain of the throughput can be explained by scattering previously introduced by the damaged antireflective coating. The shortest wavelengths are more affected by scattering and this is exactly what we see.

We have also measured the precision in the relative transmission between two stars as a function of time. For this, we observed over about an hour (so that the rotator moved by more than 20 degrees) a given field of stars. Observations were done in clear conditions in the V-band filter and the exposure time was 10s. The magnitudes of some of the brightest non-saturated stars in the field were then measured for each frame, and we looked at the dispersion in the obtained light curves of the stars (where we removed the sky variations by subtracting the mean light curve of all stars, and ignored saturated stars). The fluxes of the selected stars were about 170,000 – 500,000 ADU. Sky flats were obtained close in time to the observations and the data were reduced using the ESO FORS pipeline, with *sextractor* being used for the photometry.

Before the LADC exchange, on the night of 2014-10-28, we monitored the field around the standard star PG0231 for 50 min during which the rotator moved about 25 deg. The airmass of the field was 1. After the LADC exchange, on the night of 2014-11-13, we monitored the standard star field around NGC 2298 for 78 min during which the rotator moved 47 deg. The airmass was 1.09.

Figure 5 shows that the dispersion of the points is clearly smaller after the exchange compared to before. The standard deviation of the light curves for a star with a S/N ~ 750 decreased from 3 mmag before the exchange to 1.9 mmag after the exchange. This is, however, still larger than what we expect from the white noise, so there could still be some systematics in the data[2].

Back in business

The fact that the dispersion of points has decreased after the LADC exchange suggests that the removal of the coating is beneficial and should allow to again study exoplanet transits. However, this could only be checked on a real case, so as to determinate exactly the achieved precision. Thus, on the night of 15 to 16 November 2014, we observed the transit of WASP-19b between 2014-11-16 05:16 UT and 08:49 UT (Prog. ID: 60.A-9203(F); the data are publicly available in the ESO science archive) under thin cirrus, and with the LADC parked and in simulation as it is usually done for such observations. WASP-19b was chosen as previous observations done with FORS2 existed (taken on 16 April 2012) and would thus allow a direct comparison, while the transit duration of WASP-19b is also very short (1h32) and it is thus possible to cover it without needing too much time. Observations were done with the MXU, with 10" wide slits placed on several comparison stars, in the same configuration as for the 2012 observations. The grism 600RI (with the order sorter filter GG435) was used and the data were binned at a final 20 nm spectral resolution.

---

[2] This would, however, need to be confirmed by a more detailed analysis, as the observations of the WASP-19b transit described below seem to indicate that the level of systematics is very small.

The observations from 2012 reveal light curves with quite complex systematics (especially in the middle of the transit) that could not be removed even with high order polynomial or extinction correction functions (see Fig. 6). On the other hand, the new observations, done after the exchange, show much smoother light curves, that can be detrended using a second-order polynomial. The final, detrended light curve allows modelling, providing the parameters of the transit with good accuracy. The post egress out-of-transit residuals in the light curve are 760 μmag, very close to the value we expect from photon noise alone. This seems to indicate that the systematics that affected FORS2 have been significantly reduced. The comparison of the planetary radius as a function of wavelength (the transmission spectrum) that we obtain with the new data and those from the literature is shown in Fig. 7, highlighting the excellent agreement (see Sedaghati et al., 2015, for a more detailed analysis). The error bars of the data set (due to the poor observing conditions and lack of suitable reference stars) do not allow us to distinguish yet between different models of the planetary atmosphere. Nevertheless, they represent the highest spectral-resolution transmission spectrum of WASP-19b and show the new potential of FORS2 to study the atmosphere of exoplanets. We hope this is thus the beginning of a new era for FORS2.

*Boffin et al., ESO Messenger, March 2015*

## Figures

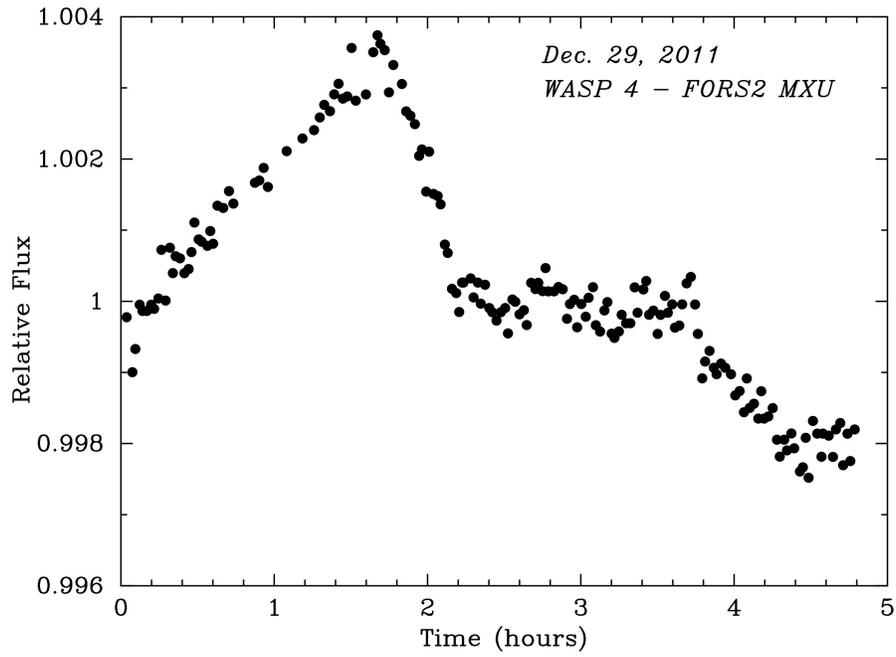

Fig. 1: Differential Light curve of the source WASP 4 obtained in the z-band using FORS2 in MXU mode in Dec. 2011. Most of the variations seen in this plot are attributed to variations in throughput within the FORS2 LADC.

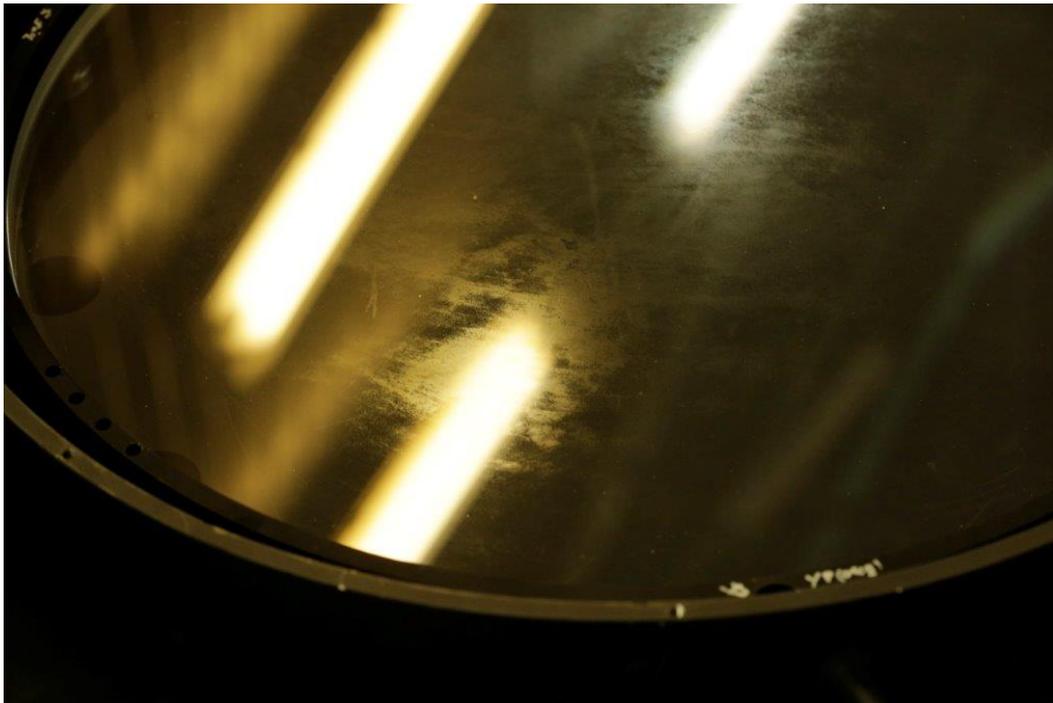

Fig. 2: Photo of the FORS2 LADC prism (with coating) after removal. The wearing off of the coating in various places is obvious. Note that the bright regions are reflections of the neon lights on the ceiling.

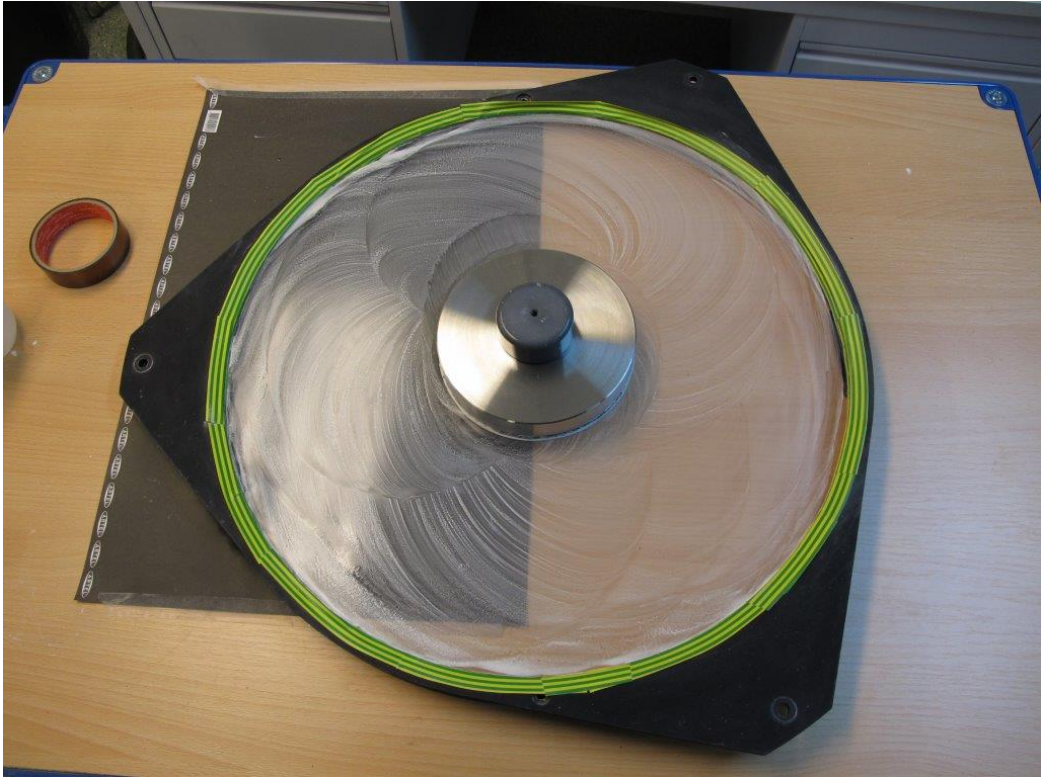

Fig. 3: Removal of the coating on the FORS1 LADC prisms.

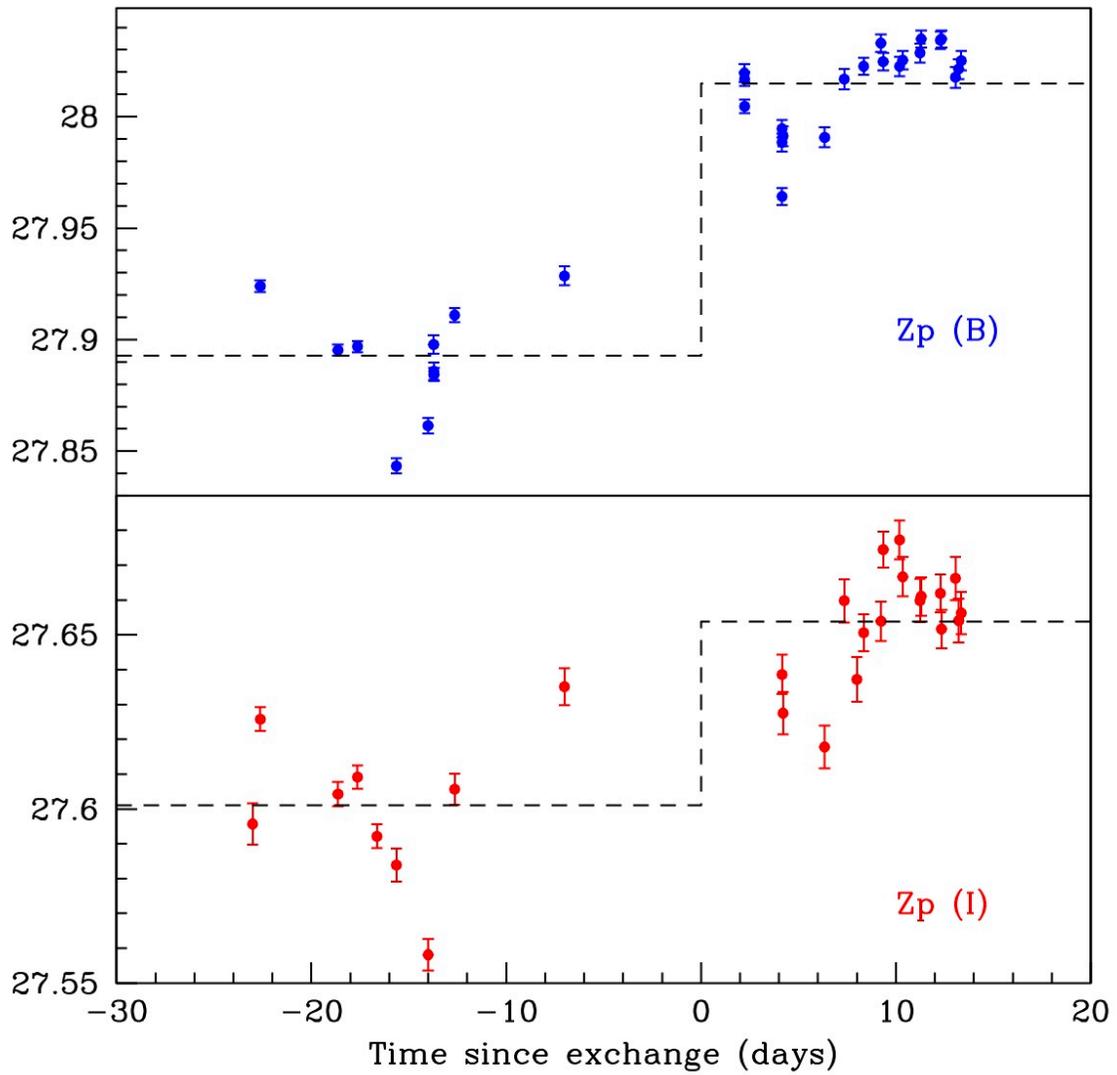

Fig. 4: FORS2 zero points measured before and after the exchange of the prisms of the LADC in the B-band (top) and in the I-band (bottom). The figures clearly show the improvement in the zero points. The dashed curve indicates the mean values and highlight this improvement.

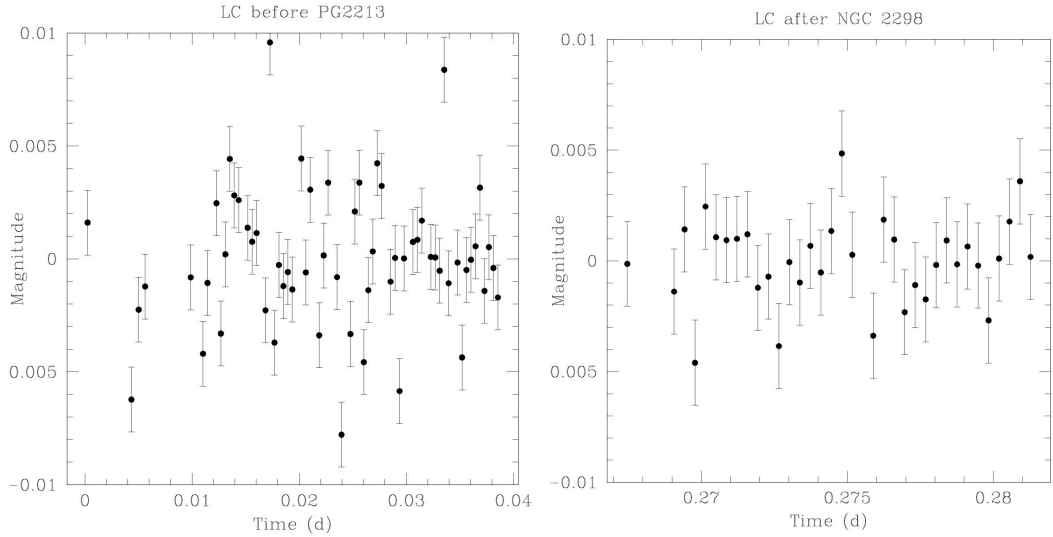

Fig. 5: Light curves in the V-band obtained before (left) and after (right) the LADC prisms exchange for a relatively bright star with a S/N~750.

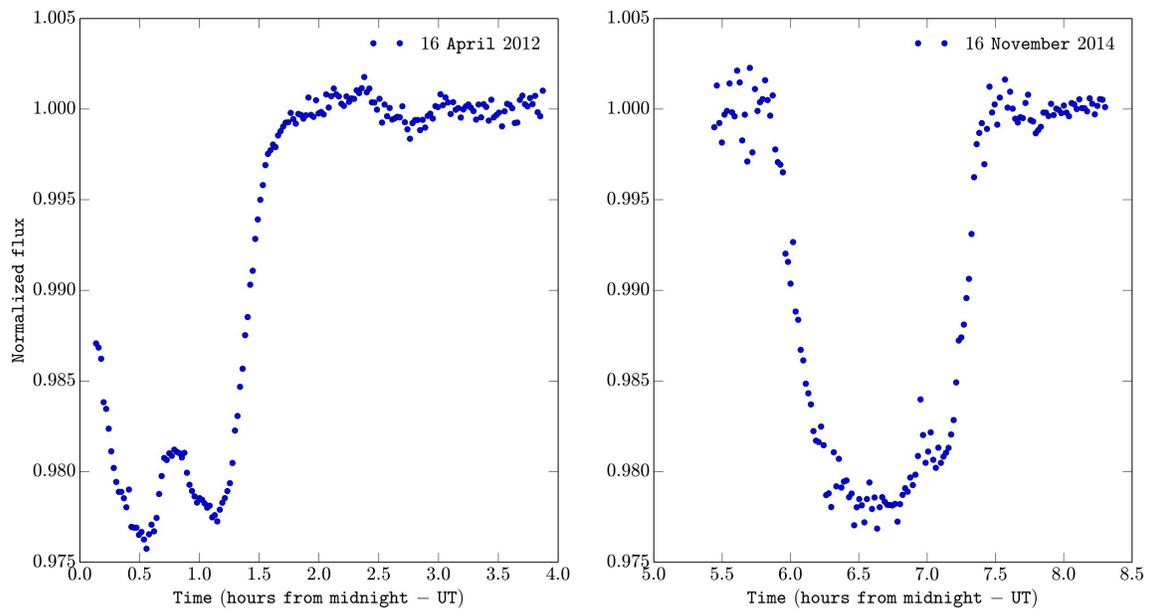

Fig. 6: Light curve of WASP-19 obtained in April 2012, i.e. before the LADC prisms exchange (left) and in November 2014, after the exchange (right), using the 600RI grism and integrating the spectra over the full wavelength domain ("white light"). Large systematics in the middle of the transit in 2012 are clearly visible.

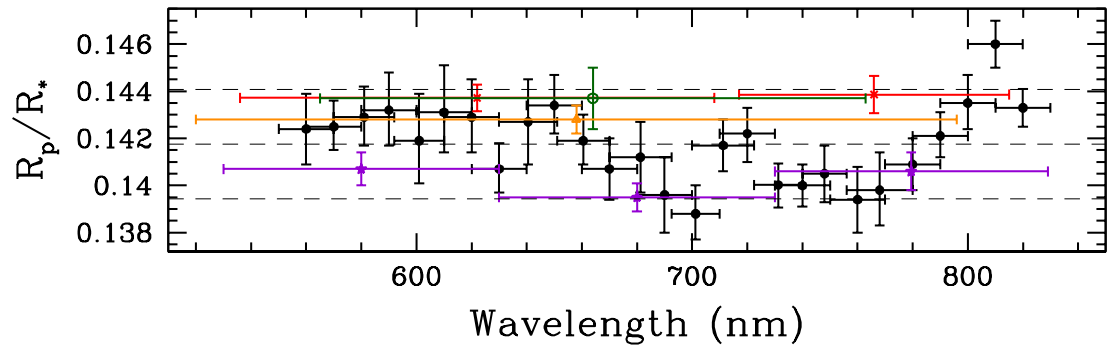

Fig. 7: Transmission spectrum of WASP-19b based on our FORS2 data (with grism 600RI and 20 nm bin) from November 2014 (black, filled dots), compared to values from the literature (colour points). The vertical error bars represent the errors in the fractional radius determination, while the horizontal bars are the FWHM of the passbands used. Note the high spectral resolution of the FORS2 data, compared to what was available until now. The dashed lines represent the weighted mean, and the interval of plus or minus three scale heights.